\documentclass[11pt,a4paper]{article}

\usepackage[a4paper,textwidth=14cm,top=2.5cm,bottom=2.5cm]{geometry}
\usepackage{amsmath,amssymb}
\usepackage{graphicx}
\usepackage{bm}
\usepackage{booktabs}
\usepackage{placeins}
\usepackage[numbers,sort&compress]{natbib}
\usepackage{microtype}
\usepackage[colorlinks=true,citecolor=blue,linkcolor=blue,urlcolor=blue]{hyperref}

\begin{document}

\title{Cartesian tensor equivariant machine-learning force field for spin-dependent atomistic simulations}

\author{%
Junjie Wang\textsuperscript{1,*}, Yijie Zhu\textsuperscript{1,*},
Zhongwei Zhang\textsuperscript{1}, Zhiyue Guo\textsuperscript{1},\\
Xudong Zhu\textsuperscript{2}, Lixin He\textsuperscript{2,3,4},
Chi Ding\textsuperscript{1}, and Jian Sun\textsuperscript{1,\(\dagger\)}\\[0.6em]
\small \textsuperscript{1}National Laboratory of Solid State Microstructures,\\
\small School of Physics and Collaborative Innovation Center of Advanced Microstructures,\\
\small Nanjing University, Nanjing 210093, China\\
\small \textsuperscript{2}Institute of Artificial Intelligence, Hefei Comprehensive National Science Center,\\
\small Hefei 230088, China\\
\small \textsuperscript{3}Laboratory of Quantum Information, University of Science and Technology of China,\\
\small Hefei 230026, China\\
\small \textsuperscript{4}Hefei National Laboratory, University of Science and Technology of China,\\
\small Hefei, Anhui 230088, China\\[0.4em]
\small \textsuperscript{*}These authors contributed equally.\quad
\textsuperscript{\(\dagger\)}Corresponding author: \href{mailto:jiansun@nju.edu.cn}{jiansun@nju.edu.cn}}

\date{}

\maketitle

\begin{abstract}
Magnetic materials exhibit an intricate coupling between atomic structure and spin degrees of freedom, posing a fundamental challenge for atomistic simulations across experimentally relevant length and time scales. Here we introduce HotPP-Spin, a spin-dependent extension of HotPP for magnetic machine learning interatomic potentials, built on Cartesian tensor equivariant message passing. Atomic magnetic moments are treated as explicit axial-vector degrees of freedom, while spatial-inversion and time-reversal parities are propagated through the tensor couplings. This construction provides a unified representation of exchange-dominated and spin-orbit-induced interactions without imposing predefined analytical interaction forms. A scalar spin-dependent potential energy surface yields energy-conserving atomic forces and magnetic effective fields through differentiation. Benchmarks spanning collinear magnetism, noncollinear magnetism, and spin-orbit-coupling-induced magnetic anisotropy show that HotPP-Spin accurately describes magnetic energy landscapes, magnetic forces, and magnetic-order-dependent energy-volume relations within the same general framework. For H-phase monolayer VSe\(_2\), stochastic spin-dynamics simulations using the learned magnetic effective fields locate the finite-size magnetic ordering crossover at 415--435~K, in close numerical agreement with the reported experimental value of \(418.5\pm7.8\)~K. These results establish Cartesian tensor message passing as a general route for connecting first-principles magnetic energetics with large-scale atomistic simulations of coupled structural and spin phenomena.
\end{abstract}


\section{\label{sec:introduction}Introduction}

Magnetic materials underpin modern information storage, sensing, and emerging spintronic and quantum technologies. Despite their importance, a predictive microscopic description of magnetic ordering across structural distortions, spin-orbit coupling, and finite-temperature fluctuations remains challenging, particularly when atomic structures and spin configurations evolve together~\cite{Wu2024UltrafastDemagnetization,Weber2022SpinPhonon,Ding2024SpinPeierls}. Atomistic simulations of these phenomena therefore require energy models that resolve both types of degrees of freedom. Density-functional theory (DFT) provides first-principles magnetic energies, atomic forces, magnetic moments, and spin-orbit-induced magnetic anisotropy, but its cost limits simulations of large supercells, long time scales, and extensive thermal sampling~\cite{Nikolov2021Magnetoelastic,Rinaldi2024NoncollinearACEFe}.

Machine learning interatomic potentials (MLIPs) extend first-principles accuracy to larger atomistic simulations~\cite{Unke2021MLFF}. Conventional MLIPs generally learn an energy from atomic positions and chemical species, but this description cannot distinguish spin configurations with the same atomic structure. A magnetic MLIP should instead learn a spin-dependent potential energy surface,
\begin{equation}
E = E(\{R_i,Z_i,S_i\}),
\label{eq:magnetic_pes_intro}
\end{equation}
where \(S_i\) denotes the spin or local magnetic moment associated with atom \(i\).

Early machine-learning approaches retained explicit spin-Hamiltonian structures. Yu et al. supplemented a Heisenberg term with a nonlinear artificial-neural-network spin potential, increasing flexibility while treating spin degrees of freedom on a fixed lattice~\cite{Yu2022ComplexSpinHamiltonian}. Nikolov et al. coupled a fitted spin Hamiltonian to a spectral neighbor analysis potential, thereby constructing a data-driven magnetoelastic potential energy surface for spin-lattice dynamics while retaining an explicit decomposition of lattice and spin contributions~\cite{Nikolov2021Magnetoelastic}. Magnetic moment tensor potentials and magnetic high-dimensional neural network potentials subsequently introduced scalar magnetic variables into established interatomic-potential frameworks, with magnetic-force training improving relaxation and equilibration~\cite{Novikov2022MagneticMTP,Eckhoff2021MagneticHDNNP,Kotykhov2024MagneticForcesMTP}. Such scalar-spin representations are primarily suited to collinear magnetism. DeepSPIN, spectral vector-field descriptors, and magnetic atomic cluster expansion instead encode vector magnetic moments, extending the description to noncollinear spin-lattice systems~\cite{Yang2024DeepSPIN,Domina2022SpectralNeighborSpin,Drautz2020MagneticACE,Rinaldi2024NoncollinearACEFe}.

Equivariant message passing neural networks (MPNNs) based on spherical harmonics and irreducible tensor products, including NequIP, MACE, and Allegro, have achieved high accuracy for nonmagnetic interatomic potentials~\cite{Batzner2022NequIP,Batatia2022MACE,Musaelian2023Allegro}. Extending equivariant message passing to magnetic systems provides a systematic tensor representation of directional spin and structural variables. SpinGNN represents spin interactions through scalar spin products and distance-dependent spin-lattice terms~\cite{Yu2024SpinGNN}. SpinGNN++ uses time-reversal-equivariant convolutions for higher-order spin-lattice interactions, whereas MagNet embeds magnetic-moment vectors in spatially equivariant tensor products and is trained on SOC-inclusive noncollinear data~\cite{Yu2024TENN,Yuan2024MagNet}. More recently, mMACE extended higher-order equivariant MACE message passing with vector magnetic moments for collinear and noncollinear systems, with and without SOC~\cite{Ho2026mMACE}.

Here we introduce HotPP-Spin, the spin-dependent extension of the nonmagnetic Cartesian tensor message-passing architecture HotPP~\cite{Wang2024HotPP}. HotPP-Spin treats atomic magnetic moments as explicit axial-vector inputs and couples them to polar structural vectors through Cartesian tensor operations. The tensor channels track their distinct spatial-inversion parity and the time-reversal parity under \(S_i\rightarrow-S_i\), while the scalar energy remains invariant. Without explicit spin-orbit coupling (SOC), the construction represents exchange-dominated interactions governed by relative spin orientations; with SOC, it can represent the dependence of energy on spin orientation relative to the lattice. HotPP-Spin thus extends Cartesian tensor message passing to spin-dependent potential energy surfaces without prescribing analytical interaction forms. Atomic forces and magnetic effective fields follow from differentiation of the learned energy with respect to atomic positions and magnetic moments, respectively. We assess the framework across collinear and noncollinear magnetic systems, with and without SOC, through alloy energetics, magnetic-order-dependent energy--volume relations, spin-orbit-induced magnetic anisotropy, and finite-temperature spin dynamics. These results demonstrate the applicability of Cartesian tensor message passing to energy-conserving atomistic simulations with coupled lattice and spin degrees of freedom.

\section{\label{sec:method}Materials and method}

\subsection{\label{sec:method_symmetry}Symmetry requirements}

The central symmetry distinction in a magnetic MLIP is between structural vectors and spin vectors.
For an orthogonal spatial transformation \(Q\), a bond vector is a polar vector and transforms as
\begin{equation}
  r_{ij} \rightarrow Q r_{ij},
  \label{eq:polar_transform}
\end{equation}
whereas a spin vector is an axial vector and transforms as
\begin{equation}
  S_i \rightarrow \det(Q) Q S_i .
  \label{eq:axial_transform}
\end{equation}
Thus, under spatial inversion \(Q=-I\), \(r_{ij}\rightarrow -r_{ij}\), while \(S_i\rightarrow S_i\).
Under time reversal, the structural variables are unchanged and the spins change sign,
\begin{equation}
  R_i\rightarrow R_i,\qquad Z_i\rightarrow Z_i,\qquad S_i\rightarrow -S_i .
  \label{eq:time_reversal}
\end{equation}
In the absence of external time-reversal-breaking fields, the total energy must be even under Eq.~\eqref{eq:time_reversal}.

HotPP-Spin labels each HotPP-style Cartesian hidden feature by its tensor rank and two parities.
A feature \({}^{(\ell,p,\tau)}\mathbf h_i^t\) at message-passing layer \(t\) has rank \(\ell\), full spatial-inversion parity \(p=\pm 1\), and time-reversal parity \(\tau=\pm 1\).
Under a general orthogonal transformation,
\begin{equation}
  {}^{(\ell,p,\tau)}\mathbf h_i^t
  \rightarrow
  \left[p(-1)^\ell\right]^{(1-\det Q)/2} Q^{\otimes \ell}
  {}^{(\ell,p,\tau)}\mathbf h_i^t,
  \label{eq:tensor_o3_transform}
\end{equation}
and under time reversal,
\begin{equation}
  {}^{(\ell,p,\tau)}\mathbf h_i^t
  \rightarrow
  \tau\,{}^{(\ell,p,\tau)}\mathbf h_i^t.
  \label{eq:tensor_time_transform}
\end{equation}
For pure inversion \(Q=-I\), Eq.~\eqref{eq:tensor_o3_transform} reduces directly to \({}^{(\ell,p,\tau)}\mathbf h_i^t\rightarrow p\,{}^{(\ell,p,\tau)}\mathbf h_i^t\), so \(p\) is the complete inversion eigenvalue.
Structural tensors generated from bond vectors are time-even, whereas every occurrence of a spin vector changes the time-reversal parity.
Only scalar channels with rank \(\ell=0\), even spatial parity, and even time-reversal parity are allowed to contribute directly to the total energy.
This restriction applies only to the final scalar readout.
Intermediate features retain all symmetry-allowed \((\ell,p,\tau)\) sectors, including spatial- or time-reversal-odd channels, which may contribute through couplings whose final output is \((0,+,+)\).
It therefore equates configurations related by inversion of the complete lattice--spin system, but does not require opposite spin chiralities on the same inversion-asymmetric lattice to be degenerate.
This construction preserves rotational equivariance for intermediate tensors while guaranteeing rotational, inversion, and time-reversal invariance of the final energy.

\subsection{\label{sec:method_pes}Overall network architecture}

HotPP-Spin extends the HotPP Cartesian tensor message-passing framework~\cite{Wang2024HotPP} from a structural potential to a spin-dependent magnetic potential.
For a periodic configuration, the input consists of atomic positions \(R_i\), chemical species \(Z_i\), lattice vectors \(L\), and spin vectors \(S_i\), where \(i=1,\ldots,N\).
The model learns a scalar spin-dependent potential energy surface,
\begin{equation}
  E_\theta = E_\theta(\{R_i,Z_i,S_i\},L),
  \label{eq:hotspot_energy}
\end{equation}
with parameters \(\theta\).
The total energy is written as a sum of atom-centered contributions,
\begin{equation}
  E_\theta = \sum_i \varepsilon_i,
  \label{eq:atomic_energy_sum}
\end{equation}
where each \(\varepsilon_i\) depends on the local chemical, structural, and magnetic environment within a finite cutoff.
Periodic boundary conditions enter through neighbor vectors
\begin{equation}
  r_{ij\bm n}=R_j+\bm n L-R_i ,
  \label{eq:neighbor_vector}
\end{equation}
with lattice image \(\bm n\).
Using relative vectors enforces translational invariance, and using shared local functions and neighbor summations enforces permutation symmetry among atoms of the same species.

\begin{figure}[!t]
\centering
\includegraphics[width=\textwidth]{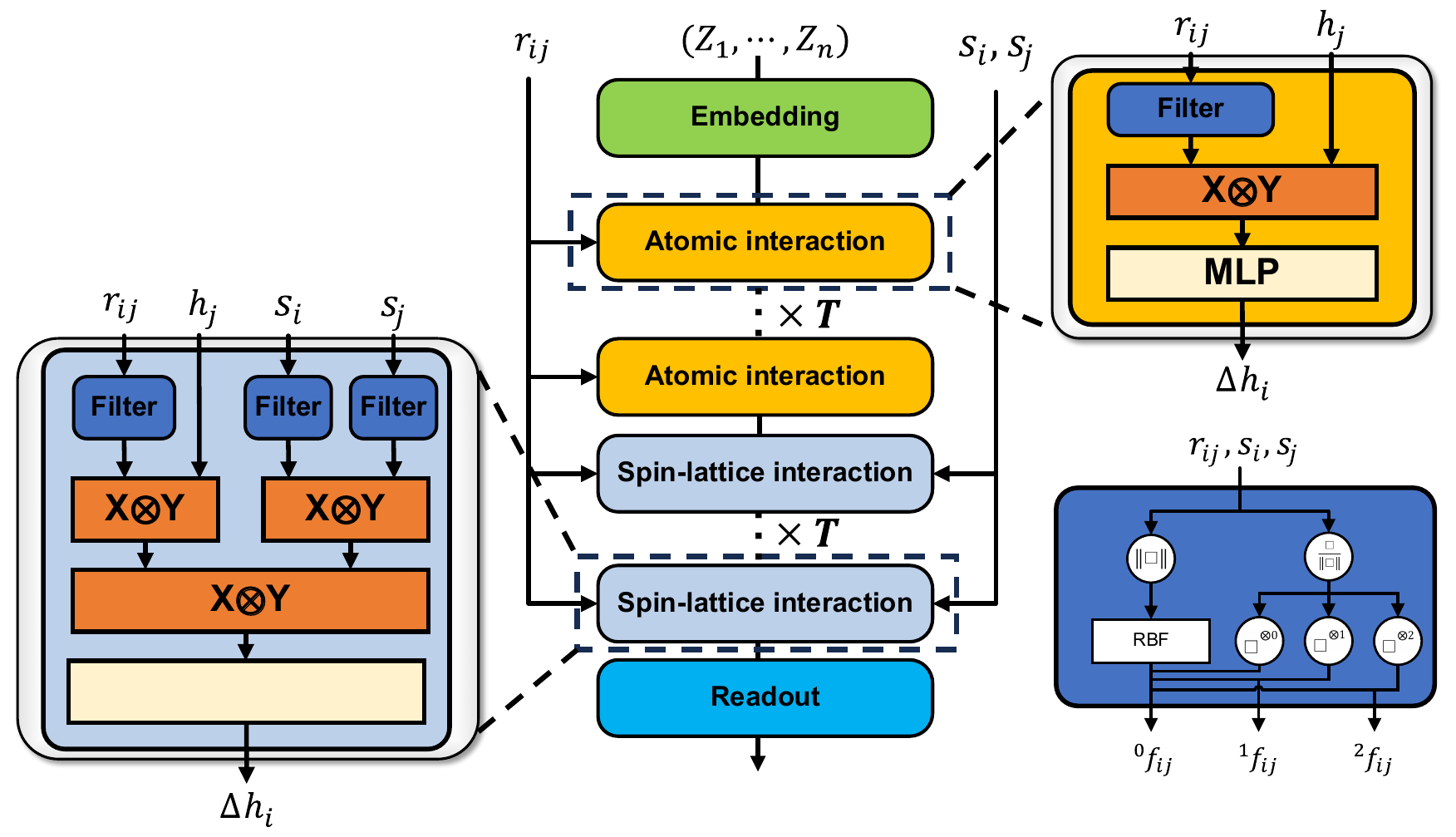}
\caption{\label{fig:hotspot_architecture}
Schematic architecture of HotPP-Spin.
Atomic positions, chemical species, lattice information, and spin vectors are embedded into tensor features and passed through interaction blocks that update structural and spin-lattice channels.
Atomic interaction blocks propagate structural information such as neighbor vectors and hidden features, while spin-lattice interaction blocks couple structural features with spin vectors.
When SOC is disabled, the two upper orange tensor-product modules directly connected to \(S_i\) and \(S_j\) in the left panel are restricted to scalar outputs with \(\ell_{\mathrm o}=0\).
Repeated interaction blocks are followed by a scalar readout, from which atomic forces and magnetic effective fields are obtained by differentiating the learned energy.}
\end{figure}

Figure~\ref{fig:hotspot_architecture} summarizes the HotPP-Spin architecture and the interaction blocks used to propagate structural and spin information. The spin variables in Eq.~\eqref{eq:hotspot_energy} may represent local magnetic moments or normalized spin directions, depending on the reference data.
HotPP-Spin does not reduce this information to a scalar magnetic magnitude.
Instead, each \(S_i\) is treated as a vector degree of freedom so that collinear and noncollinear spin configurations with the same atomic positions can be assigned different energies and derivatives.
HotPP-Spin follows the Cartesian tensor strategy of HotPP rather than using spherical harmonics as the primary representation.
Neighbor information is first embedded through species-dependent radial functions of \(|r_{ij}|\) and Cartesian tensor products of the unit or displacement vector.
Spin information is introduced through axial-vector channels associated with \(S_i\) and neighboring \(S_j\).
Messages are then formed by tensor products, contractions, and nonlinear mixing among structural, spin, and hidden tensor features, with the parity labels in Eqs.~\eqref{eq:tensor_o3_transform} and \eqref{eq:tensor_time_transform} updated at each coupling step.

For Cartesian tensors \(\mathbf X\) and \(\mathbf Y\) of ranks \(\ell_X\) and \(\ell_Y\), respectively, the \(X\otimes Y\) module uses the HotPP tensor contraction~\cite{Wang2024HotPP}
\begin{equation}
\begin{split}
  &\left[\mathbf X\otimes_z\mathbf Y\right]_{
  a_1\cdots a_{\ell_X-z}b_1\cdots b_{\ell_Y-z}}
  \\
  &\quad=
  X_{a_1\cdots a_{\ell_X-z}c_1\cdots c_z}
  Y_{b_1\cdots b_{\ell_Y-z}c_1\cdots c_z},
  \\
  &(\ell_{\mathrm o},p_{\mathrm o},\tau_{\mathrm o})
  =(\ell_X+\ell_Y-2z,\,p_Xp_Y,\,\tau_X\tau_Y),
\end{split}
\label{eq:hotspot_tensor_coupling}
\end{equation}
where \(0\le z\le\min(\ell_X,\ell_Y)\), and \(z=0\) gives the outer product.
HotPP-Spin supplements the original rank rule with spatial and time-reversal parities \((p,\tau)\).
For example, in the \((\ell,p,\tau)\) notation, a polar bond vector belongs to \((1,-,+)\), whereas an axial spin vector belongs to \((1,+,-)\).
Consequently,
\begin{equation}
\begin{gathered}
  S_i\!\cdot\!S_j\sim(0,+,+),
  \qquad
  r_{ij}\!\cdot\!S_i\sim(0,-,-),
  \\
  (r_{ij}\!\cdot\!S_i)
  (r_{ij}\!\cdot\!S_j)\sim(0,+,+).
\end{gathered}
\end{equation}
The spin--spin scalar can contribute directly to the energy, whereas the single bond--spin contraction cannot.
Coupling two such bond--spin odd scalars produces a symmetry-allowed structural--spin anisotropic scalar.
For example, \((r_{ik}\!\times\!r_{jk})\!\cdot\!(S_i\!\times\!S_j)\) is even under both spatial inversion and time reversal, yet changes sign when the spin chirality is reversed while the lattice is held fixed.
These parity rules determine whether a coupling is symmetry-allowed; its physical presence and magnitude are learned from the reference data.
Spin--spin contractions describe exchange-dominated changes associated with relative spin orientations, whereas mixed structural--spin contractions permit orientation dependence relative to the lattice when such dependence is present in the reference data.

Following the edge construction of HotPP~\cite{Wang2024HotPP}, the spin--lattice block in Fig.~\ref{fig:hotspot_architecture} can be summarized as
\begin{equation}
\begin{split}
  \bar r_{ij\bm n}^{t,\ell}&=
  \mathrm{Filter}_r^{t,\ell}(r_{ij\bm n})
  =f_\ell^t(d_{ij\bm n})u_{ij\bm n}^{\otimes\ell},
  \\
  \bar S_a^{t,k}&=
  \mathrm{Filter}_s^{t,k}(S_a)
  =g_k^t(|S_a|)\widehat S_a^{\otimes k},\quad a\in\{i,j\},
  \\
  \bar S_{ij}^{t,\ell_s}&=
  \left[\bar S_i^{t,k_i}\otimes
  \bar S_j^{t,k_j}\right]_{\ell_s},
  \\
  \Delta h_i^t&=
  \sum_{(j,\bm n)\in\mathcal N(i)}
  \mathrm{MLP}_t\!\left\{
  \left(\bar r_{ij\bm n}^{t,\ell}\otimes h_j^t\right)
  \otimes
  \bar S_{ij}^{t,\ell_s}
  \right\},
  \\
  h_i^{t+1}&=h_i^t+\Delta h_i^t .
\end{split}
\label{eq:message_passing}
\end{equation}
Here \(d_{ij\bm n}=|r_{ij\bm n}|\), \(u_{ij\bm n}=r_{ij\bm n}/d_{ij\bm n}\), and \(\widehat S_a=S_a/|S_a|\); species dependence of the learnable radial functions is left implicit.
Each \(\otimes\) collects the symmetry-allowed contractions in Eq.~\eqref{eq:hotspot_tensor_coupling}, while channel mixing and tensor activation are absorbed into \(\mathrm{MLP}_t\).
Without SOC, the spin--spin branch is restricted to \(\ell_s^{\max}=0\).
For equal input rank \(k\), its scalar channel is
\begin{equation}
  \bar S_{ij}^{t,0}
  =g_k^t(|S_i|)g_k^t(|S_j|)
  (\widehat S_i\!\cdot\!\widehat S_j)^k,
  \label{eq:nosoc_spin_scalar}
\end{equation}
so only spin magnitudes, \(S_i\!\cdot\!S_j\), and their higher-order scalar combinations enter the structural message.
These quantities are invariant under a simultaneous rotation of all spins relative to a fixed lattice and therefore describe exchange-like interactions without lattice-locked SOC terms.
With SOC, symmetry-allowed channels with \(\ell_s>0\) are retained and subsequently coupled to \(\bar r_{ij\bm n}^{t,\ell}\otimes h_j^t\), allowing the energy to depend on spin orientation relative to the lattice.
The actual cutoff, channel dimensions, tensor ranks, and number of message-passing layers are dataset-dependent implementation parameters, given for each benchmark in Appendix~\ref{app:computational_details}.

Because HotPP-Spin is an energy-conserving model, atomic and magnetic responses are obtained by differentiating Eq.~\eqref{eq:hotspot_energy}.
Atomic forces are defined as
\begin{equation}
  F_i = -\frac{\partial E_\theta}{\partial R_i}.
  \label{eq:atomic_force}
\end{equation}
The spin derivative defines the magnetic force, or equivalently the negative energy gradient in spin space,
\begin{equation}
  B_i^{\mathrm{ML}} =
  -\frac{\partial E_\theta}{\partial S_i}.
  \label{eq:magnetic_force}
\end{equation}
This quantity is the training target when magnetic-force labels are available.
For fixed-length spin dynamics, only the component perpendicular to \(S_i\) changes its orientation; the radial component does not contribute because \(S_i\times B_i^{\mathrm{ML}}=S_i\times B_{i,\perp}^{\mathrm{ML}}\).
The LLG equation therefore uses the learned magnetic force directly rather than a separately trained torque.

The model is trained against first-principles reference energies and, when available, atomic forces, magnetic forces, and virials.
A representative mean-squared-error objective is
\begin{equation}
\begin{split}
  \mathcal L
  ={}&
  w_E\,\mathrm{MSE}\!\left(
    \frac{E_\theta}{N},\frac{E_{\mathrm{ref}}}{N}
  \right)
  +w_F\,\mathrm{MSE}\!\left(F_\theta,F_{\mathrm{ref}}\right)
  \\
  &+
  w_B\,\mathrm{MSE}\!\left(B_\theta^{\mathrm{ML}},B_{\mathrm{ref}}\right)
  +w_{\mathcal V}\,\mathrm{MSE}\!\left(
    \frac{\mathcal V_\theta}{N},
    \frac{\mathcal V_{\mathrm{ref}}}{N}
  \right).
\end{split}
\label{eq:training_loss}
\end{equation}
Here \(N\) is the number of atoms, \(\mathcal V\) is the virial tensor, and each MSE is averaged over the corresponding configurations and Cartesian components.
The weights \(w_E\), \(w_F\), \(w_B\), and \(w_{\mathcal V}\) control the relative balance among the training targets.
This derivative-based training is essential for spin and spin-lattice dynamics because the same learned scalar energy must generate both atomic and magnetic forces.

\subsection{\label{sec:method_soc}SOC module and time-reversal parity}

The symmetry labels introduced in Sec.~\ref{sec:method_symmetry} track the spatial-inversion and time-reversal characters of every tensor channel.
Together with the spin-rank restriction described in Eqs.~\eqref{eq:message_passing} and \eqref{eq:nosoc_spin_scalar}, they distinguish exchange-like scalar channels in the no-SOC construction from the \(\ell_s>0\) spin--lattice channels that enable SOC-induced magnetic anisotropy.
In the parity-resolved SOC model, channels with different \((p,\tau)\) labels remain separate throughout the network, and the energy readout is restricted to \((\ell,p,\tau)=(0,+,+)\).
No such restriction is imposed on the intermediate tensor features.
The parity-unresolved SOC model used below is an ablation that retains tensor rank \(\ell\) but merges channels with different spatial and time-reversal parities.
Its rank-zero readout is therefore not guaranteed to belong to \((0,+,+)\) and can acquire spurious responses to spatial inversion or time reversal.
This ablation is used only to test the role of explicit parity resolution, not as a model for the material benchmarks.
These architectural constraints determine which dependencies are permitted, while their magnitudes are learned from the reference data.
SOC-sensitive behavior is assessed using the SOC-inclusive H-phase monolayer VSe\(_2\) data; the FeAl and Fe benchmarks instead test collinear multi-component fitting and noncollinear exchange-dominated behavior, respectively.

\begin{figure}[!t]
\centering
\includegraphics[width=\textwidth,height=0.12\textheight,keepaspectratio]{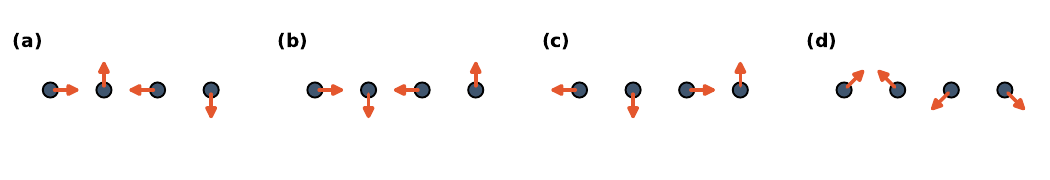}
\caption{\label{fig:hotpp_chirality_toy}
(a) Reference \(0^\circ,90^\circ,180^\circ,270^\circ\) spin sequence and the configurations obtained by (b) reversing spin chirality, (c) time reversal, and (d) a \(45^\circ\) in-plane global spin rotation.
The low-symmetry nonmagnetic environment used in the architecture audit is included in the model input but omitted from the schematic.}
\end{figure}

\begin{table}[!t]
\small
\caption{\label{tab:hotpp_chirality_toy}Energy responses of untrained networks under the operations illustrated in Fig.~\ref{fig:hotpp_chirality_toy}. ``Invariant'' denotes no energy change within single-precision numerical accuracy, whereas ``Changes'' denotes a finite response across random initializations and configurations. All responses are assessed against the expected behavior of an SOC-active low-symmetry system; \(\checkmark\) and \(\times\) denote agreement and disagreement, respectively. The parity-unresolved SOC model is an ablation used only to test symmetry enforcement.}
\centering
\resizebox{\textwidth}{!}{%
\begin{tabular}{lccc}
\toprule
Model & Spin-chirality reversal & Time reversal & In-plane spin rotation \(45^\circ\) \\
\midrule
Model without SOC capability & Invariant (\(\times\)) & Invariant (\(\checkmark\)) & Invariant (\(\times\)) \\
SOC, parity-unresolved & Changes (\(\checkmark\)) & Changes (\(\times\)) & Changes (\(\checkmark\)) \\
SOC, parity-resolved & Changes (\(\checkmark\)) & Invariant (\(\checkmark\)) & Changes (\(\checkmark\)) \\
\bottomrule
\end{tabular}
}
\end{table}

We isolate these architectural roles with an untrained-network symmetry audit, avoiding any dependence on a fitted target or training-set augmentation.
The input comprises a four-site magnetic chain embedded in a low-symmetry nonmagnetic environment; Fig.~\ref{fig:hotpp_chirality_toy} shows only the magnetic sites for clarity.
We evaluate eight random network initializations on 64 random spin and weakly perturbed low-symmetry structural configurations without optimization, and classify each transformation according to whether it produces an energy response beyond single-precision numerical residuals.
Table~\ref{tab:hotpp_chirality_toy} shows that the model without SOC capability and the parity-resolved SOC model are exactly time-reversal invariant at numerical precision, whereas the parity-unresolved SOC ablation has a finite response.
The model without SOC capability is also invariant under spin-chirality reversal and global spin rotation, and therefore cannot represent the corresponding finite responses expected for the SOC-active low-symmetry system used in this audit.
By contrast, both SOC networks permit finite chirality and spin-rotation responses because their spin channels couple to the fixed structural frame.
These finite responses demonstrate representational permission rather than a physical prediction by randomly initialized weights; the audit therefore separates symmetry guaranteed by construction from behavior that could otherwise be learned accidentally from augmented data.

\subsection{\label{sec:method_spin_dynamics}Finite-temperature spin dynamics}

For fixed-length stochastic spin dynamics, we follow standard atomistic spin-dynamics formulations~\cite{Skubic2008AtomisticSpinDynamics} and write \(S_i=\mu_i s_i\), where \(|s_i|=1\) and \(\mu_i=|S_i|\) is in units of \(\mu_{\mathrm B}\).
At fixed \(\mu_i\), the energy-like spin force supplied to the dynamics is
\(b_i^{\mathrm{ML}}=-\partial E/\partial s_i=\mu_i B_i^{\mathrm{ML}}\), with \(B_i^{\mathrm{ML}}\) defined in Eq.~\eqref{eq:magnetic_force}.
The corresponding tesla-valued field is \(\mathcal B_i^{\mathrm{ML}}=b_i^{\mathrm{ML}}/(\mu_i\mu_{\mathrm B})\), and the deterministic part of the stochastic Landau--Lifshitz--Gilbert equation~\cite{Mentink2010SIBLL} is
\begin{equation}
  \frac{d s_i}{d t}
  =
  -\frac{\gamma}{1+\alpha^2}
  \left[
    s_i\times \mathcal B_i^{\mathrm{ML}}
    +
    \alpha s_i\times\left(s_i\times \mathcal B_i^{\mathrm{ML}}\right)
  \right],
  \label{eq:sllg_omega}
\end{equation}
where \(\alpha\) is the Gilbert damping parameter and \(\gamma\) is the electron gyromagnetic ratio.
The normalized magnetic order parameter is evaluated as
\begin{equation}
  m =
  \frac{
    \left|\sum_i S_i\right|
  }{
    \sum_i |S_i|
  }.
  \label{eq:vse2_order_parameter}
\end{equation}
The Stratonovich SIB integrator, thermal-noise covariance, and sampling and blocking protocols are given in Appendix~\ref{app:spin_dynamics}.

\section{\label{sec:results}Results}

Across three material systems, we train four independent models on four separate datasets within the same general framework.
The benchmarks comprise collinear FeAl, noncollinear exchange-dominated Fe, constrained collinear bcc/fcc Fe for the energy--volume diagnostic in Sec.~\ref{sec:results_fe}, and trigonal-prismatic H-phase monolayer VSe\(_2\) with SOC-inclusive reference data.
Further dataset, model, first-principles, and spin-dynamics details are provided in Appendices~\ref{app:computational_details} and~\ref{app:spin_dynamics}.

\begin{table}[!t]
\footnotesize
\caption{\label{tab:feal_error_comparison}FeAl full-dataset fitting RMSEs. All models were trained and evaluated on the complete 2632-configuration dataset; the reported values are therefore in-sample fitting errors rather than held-out test errors. Stress errors are reported in GPa. The mMTP values are from Ref.~\cite{Kotykhov2024MagneticForcesMTP}. The lowest RMSE in each column is shown in bold. A dash indicates that the quantity is not defined for that model.}
\centering
\resizebox{\textwidth}{!}{%
\begin{tabular}{lcccc}
\toprule
Model & \(E\) (meV/atom) & \(F\) (meV/\AA{}) & \(\sigma\) (GPa) & \(B\) (meV/\(\mu_{\mathrm B}\)) \\
\midrule
mMTP (\(N_\psi=4\), \(w_t=0.1\))~\cite{Kotykhov2024MagneticForcesMTP} & 1.61 & 63.0 & 0.450 & 15.4 \\
HotPP (no spin) & 13.4 & 36.1 & 2.70 & -- \\
HotPP-Spin & \textbf{0.137} & \textbf{0.945} & \textbf{0.133} & \textbf{0.393} \\
\bottomrule
\end{tabular}
}
\end{table}

\subsection{\label{sec:results_feal}FeAl: collinear multi-component magnetic benchmark}

We first test HotPP-Spin on the complete 2632-configuration collinear bcc Fe-Al alloy benchmark~\cite{Kotykhov2023ConstrainedDFTFeAl,Kotykhov2024MagneticForcesMTP}.
The Fe and Al magnetic-moment magnitudes span 0--3.72 and 0--0.23~\(\mu_{\mathrm B}\), respectively.
This benchmark probes whether the model can learn energy, atomic-force, stress, and magnetic-force labels in a multi-component magnetic alloy without explicit SOC.
For a protocol-matched comparison with the published magnetic moment tensor potential (mMTP), we train HotPP-Spin on all 2632 configurations and evaluate it on the same full dataset.
HotPP-Spin gives fitting RMSEs of 0.137~meV/atom for energy, 0.945~meV/\AA{} for atomic forces, 0.133~GPa for stress, and 0.393~meV/\(\mu_{\mathrm B}\) for magnetic forces, with the force errors computed over Cartesian components.
Table~\ref{tab:feal_error_comparison} compares these values with the mMTP errors reported on the whole training set~\cite{Kotykhov2024MagneticForcesMTP}.
The mMTP row corresponds to the reported \(N_\psi=4\), \(w_t=0.1\) model, where \(N_\psi\) is the number of magnetic basis functions used to represent the magnetic-moment dependence of the mMTP radial functions and \(w_t\) is the magnetic-force weight; HotPP-Spin uses equal nominal weights for the four fitted targets, although the numerical weights are not directly comparable across the two loss definitions.
The spin-free HotPP baseline is trained on the same complete dataset using only energy, atomic-force, and stress labels and gives fitting RMSEs of 13.4~meV/atom, 36.1~meV/\AA{}, and 2.70~GPa, respectively; magnetic-force prediction is not defined without magnetic-moment inputs.
Under this matched in-sample evaluation protocol, HotPP-Spin has lower fitting RMSEs than mMTP for all four targets and lower errors than spin-free HotPP for the three shared targets; because the same configurations are used for fitting and evaluation, this comparison assesses fitting accuracy rather than generalization to unseen configurations.

\subsection{\label{sec:results_fe}Fe: noncollinear magnetic benchmark}

\begin{figure}[!t]
\centering
\includegraphics[width=\textwidth]{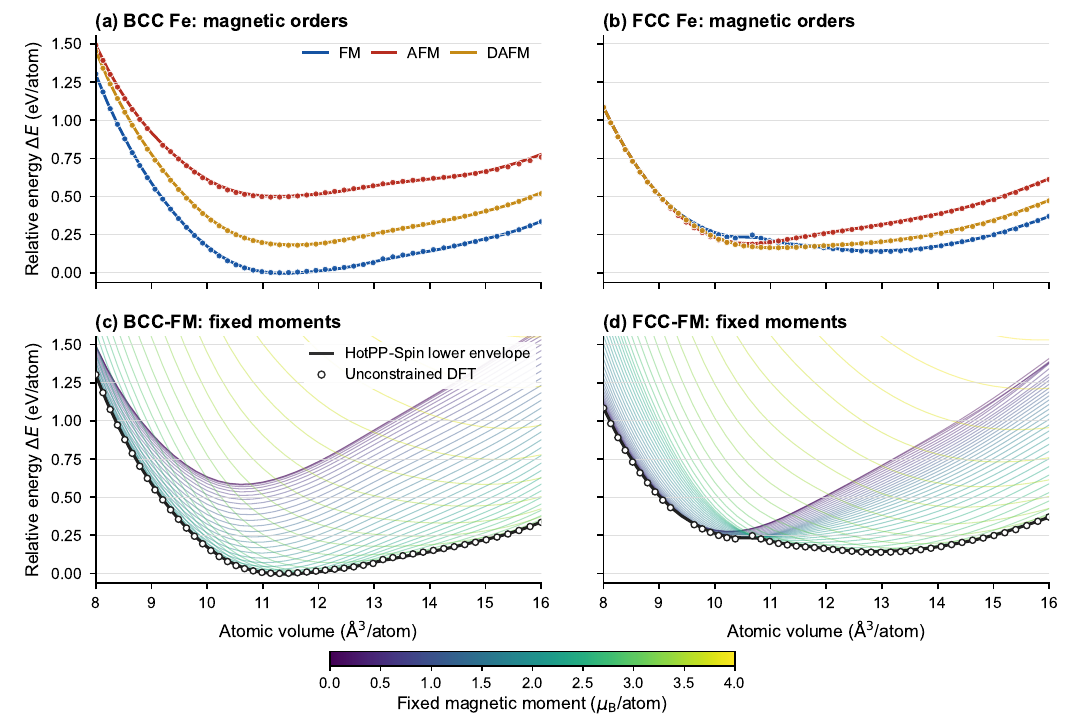}
\caption{\label{fig:fe_magnetic_orders_ev}
Energy--volume relations for ideal bcc and fcc Fe predicted by the separately trained collinear HotPP-Spin model.
(a,b) HotPP-Spin lower envelopes for the FM (blue), AFM (red), and DAFM (gold) orders, with corresponding unconstrained DFT values shown as filled circles.
(c,d) Complete BCC-FM and FCC-FM fixed-moment landscapes; all 41 predicted curves from 0 to 4~\(\mu_{\mathrm B}\) per atom in increments of 0.1~\(\mu_{\mathrm B}\) are colored by moment magnitude.
The dark curves are the HotPP-Spin lower envelopes obtained by minimizing the 41 fixed-moment predictions independently at each volume, and open circles show the unconstrained DFT results.
All energies are referenced to the minimum across the six unconstrained DFT branches; the same reference is applied to the model predictions, with no phase- or order-specific fitted offset.
The envelope is a discrete minimization over fixed-moment predictions and is not a self-consistent magnetic-moment relaxation.}
\end{figure}

We next consider noncollinear bulk Fe using the 34228 sixteen-atom configurations in the publicly released \(\mathrm{Fe}_{16}\) LCAO portion of the Fe-DeepSpin dataset.
Although the underlying first-principles calculations included SOC, the source study reported that SOC effects in pure Fe are smaller than its intrinsic model errors~\cite{Zheng2026FeDeepSpin}.
Accordingly, we train the present Fe model without explicit SOC and use the dataset as an exchange-dominated benchmark of vector magnetic degrees of freedom.
We use a fixed split of 30806 training configurations and 3422 held-out test configurations.
The magnetic-moment magnitudes span 0.100--4.271~\(\mu_{\mathrm B}\) in the training set and 0.101--4.066~\(\mu_{\mathrm B}\) in the test set, with corresponding means of 2.264 and 2.261~\(\mu_{\mathrm B}\).
On the 3422 held-out configurations, HotPP-Spin gives componentwise RMSEs of 1.207~meV/atom, 13.553~meV/\AA{}, and 6.936~meV/\(\mu_{\mathrm B}\) for energy, atomic force, and magnetic force, respectively; the source-reported DeePSPIN-DZP test RMSEs are 8.60~meV/atom, 111~meV/\AA{}, and 31.2~meV/\(\mu_{\mathrm B}\), respectively~\cite{Zheng2026FeDeepSpin}.

As a complementary collinear validation, we separately trained another model without explicit SOC on 4940 constrained ideal bcc and fcc Fe configurations spanning ferromagnetic (FM), antiferromagnetic (AFM), and double-layer antiferromagnetic (DAFM) orders; DAFM denotes successive spin layers in an \(\uparrow\uparrow\downarrow\downarrow\) sequence~\cite{Zheng2026FeDeepSpin}.
We use a fixed 4446/494 training/test split.
These configurations cover atomic volumes from 8 to 16~\AA\(^3\)/atom and magnetic-moment magnitudes from 0 to 4~\(\mu_{\mathrm B}\) per atom.
On the independent 494-configuration test subset, the dedicated collinear model gives RMSEs of 11.678~meV/atom for energy, 7.797~meV/\AA{} for atomic forces, and 14.857~meV/\(\mu_{\mathrm B}\) for magnetic forces.
Figure~\ref{fig:fe_magnetic_orders_ev} compares this separately trained model with independently calculated unconstrained DFT results.
The predicted envelopes follow the smooth DFT volume dependence and order-dependent energy separations; their RMSEs against the matched unconstrained DFT points range from 2.66 to 4.98~meV/atom across the six branches.
The unconstrained DFT trajectories are used only for this diagnostic and are absent from both the training and held-out constrained subsets.
The bcc-FM branch gives the global minimum, whereas the closer fcc branches and the volume-dependent fixed-moment envelopes expose the competition among lattice volume, magnetic order, and moment magnitude.
Comparable constrained-energy landscapes were reported for magnetic ACE and Fe-DeepSpin, likewise revealing the strong dependence of magnetic energetics on lattice volume in Fe~\cite{Rinaldi2024NoncollinearACEFe,Zheng2026FeDeepSpin}.
Magnetic ACE further used the coupled landscape in finite-temperature simulations of FM--paramagnetic and bcc--fcc transformations~\cite{Rinaldi2024NoncollinearACEFe}, while Fe-DeepSpin demonstrated a finite-temperature FM--paramagnetic transition~\cite{Zheng2026FeDeepSpin}.
Figure~\ref{fig:fe_magnetic_orders_ev} therefore validates the static energetic ingredients relevant to phase stability; transition temperatures require finite-temperature sampling and are not inferred here.

\subsection{\label{sec:results_vse2}H-phase monolayer \texorpdfstring{VSe\(_2\)}{VSe2}: spin-orbit-induced magnetic anisotropy}

We use the trigonal-prismatic H-phase monolayer, often denoted monolayer 2H-VSe\(_2\), as a two-dimensional benchmark with SOC-inclusive reference data.
The 421 training configurations comprise randomly displaced monolayer supercells and fixed-structure six-atom cells that scan the orientations of the two vanadium spins.
In the random supercells, the vanadium spin vectors have random directions and magnitudes from 0.90 to 1.10~\(\mu_{\mathrm B}\), while the selenium sites are nonmagnetic in the stored spin labels.
On the training set, the model gives RMSEs of 1.50 meV/atom for energy, 15.8 meV/\AA{} for atomic forces, and 5.08 meV/\(\mu_{\mathrm B}\) for magnetic forces.

To assess the relevant magnetic energy scales, we compare HotPP-Spin and SOC-DFT energies for in-plane (IP) and out-of-plane (OP) ferromagnetic states and an OP antiferromagnetic state in the same six-atom cell.
Per vanadium atom, HotPP-Spin gives \(E_{\mathrm{OP\text{-}FM}}-E_{\mathrm{IP\text{-}FM}}=0.567\)~meV, compared with 0.615~meV from DFT, and \(E_{\mathrm{OP\text{-}AFM}}-E_{\mathrm{OP\text{-}FM}}=132.68\)~meV, compared with 138.89~meV from DFT.
HotPP-Spin therefore reproduces both the small SOC-driven magnetic-anisotropy energy and the much larger FM--AFM energy splitting, correctly identifying the in-plane FM state as the lowest-energy configuration.
The ferromagnetic ordering preference and in-plane easy axis are consistent with previous first-principles calculations for H-phase monolayer VSe\(_2\)~\cite{Fuh2016VSe2SciRep,Sheng2021VSe2JMaterSci}.
First-principles-based spin models have yielded estimates of the Curie temperature (\(T_{\mathrm C}\)) ranging from approximately 300 to 472~K, depending on strain and computational treatment~\cite{Fuh2016VSe2SciRep,Sheng2021VSe2JMaterSci,Zhu2022AlNVSe2PRApplied}.
Experimentally, temperature-dependent magnetic-circular-dichroism measurements on 2H-VSe\(_2\) nanoflakes yielded \(T_{\mathrm C}=418.5\pm7.8\)~K~\cite{Wang2021VSe2ACSNano}.

\begin{figure}[!t]
\centering
\includegraphics[width=8.5cm]{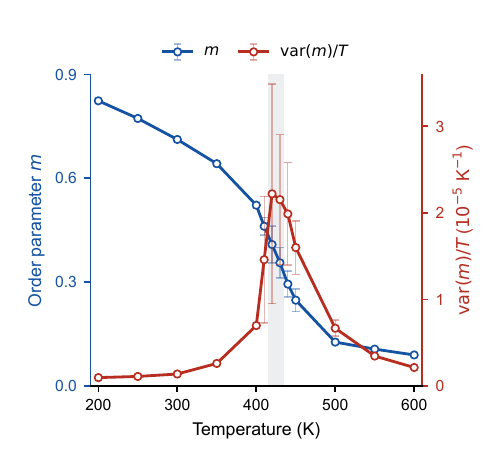}
\caption{\label{fig:vse2_spin_dynamics}
Stochastic spin dynamics for H-phase monolayer VSe\(_2\) driven by HotPP-Spin magnetic effective fields.
The normalized magnetic order parameter \(m=|\sum_i S_i|/\sum_i |S_i|\) is shown in blue against the left axis, and the fluctuation measure \(\mathrm{var}(m)/T\), computed from equilibrated trajectories, is shown in red against the right axis.
The simulations used a 2700-atom supercell; error bars show block-to-block standard deviations of the retained trajectories.
The gray shaded interval marks the approximate 415--435 K range around the fluctuation maximum and the finite-size estimate of the ordering crossover.
}
\end{figure}

To test whether the learned magnetic effective fields can be used in finite-temperature spin sampling, we performed stochastic LLG simulations for VSe\(_2\) using the dynamics described in Sec.~\ref{sec:method_spin_dynamics}.
The simulations used a 2700-atom monolayer supercell and sampled temperatures from 200 to 600 K.
As shown in Fig.~\ref{fig:vse2_spin_dynamics}, the normalized order parameter \(m\) in Eq.~\eqref{eq:vse2_order_parameter} decreases from 0.824 at 200 K to 0.522 at 400 K and 0.090 at 600 K.
For fixed site moments and the present normalization, the finite-size magnetic susceptibility satisfies \(\chi\propto N_{\mathrm{mag}}\mathrm{var}(m)/(k_{\mathrm B}T)\), so the maximum of \(\mathrm{var}(m)/T\) in the 420--430 K region provides a finite-size estimate of the magnetic ordering crossover in the HotPP-Spin-driven stochastic spin dynamics.
The resulting 415--435~K interval lies within previous theoretical estimates and is close to the experimental temperature scale for 2H-VSe\(_2\).

\FloatBarrier
\section{\label{sec:conclusion}Discussion and conclusions}

We introduced HotPP-Spin, a Cartesian tensor equivariant magnetic MLIP that treats atomic magnetic moments as explicit axial-vector degrees of freedom and propagates spatial-inversion and time-reversal parities through Cartesian tensor couplings. Differentiation of a scalar spin-dependent energy yields energy-conserving atomic forces and magnetic effective fields, providing a consistent basis for structural and spin evolution.
The FeAl and noncollinear Fe benchmarks quantify the fit quality for energies, atomic forces, and magnetic forces, while the Fe energy--volume tests reproduce magnetic-order-dependent bcc and fcc branches. For H-phase monolayer VSe\(_2\), HotPP-Spin reproduces targeted SOC-sensitive magnetic anisotropy and FM-AFM energy differences, and the learned magnetic effective fields yield a finite-size ordering crossover of 415--435~K. This range is close to the MCD-derived \(T_{\mathrm C}=418.5\pm7.8\)~K reported for a 16.9-nm-thick 2H-VSe\(_2\) flake, despite the different thickness and simulation protocol~\cite{Wang2021VSe2ACSNano}.
Looking ahead, embedding HotPP-Spin in evolutionary crystal-structure-search frameworks such as MAGUS and its symmetry-guided MAGUS 2.0 extension could enable joint exploration of atomic and magnetic configurational spaces~\cite{Wang2023MAGUS,Han2025MAGUS2}. Adapting large-cell active-learning workflows such as NEPMaker to spin-dependent descriptors and magnetic-force labels could further automate the construction of robust magnetic datasets~\cite{Wang2026NEPMaker}.

\section*{Code availability}
The HotPP-Spin implementation developed in this work is available as part of the HotPP project at \url{https://gitlab.com/bigd4/hotpp}, the spin-dynamics code is available through the PyDyn project at \url{https://gitlab.com/bigd4/pydyn}, and the ActiveMiao active-learning module is available at \url{https://gitlab.com/bigd4/activemiao}.

\section*{Acknowledgements}
This work was supported by the National Natural Science Foundation of China (Grant Nos.~12125404, T2495231, and 12504277); the National Key R\&D Program of China (Grant No.~2022YFA1403201); the Advanced Materials--National Science and Technology Major Project (Grant No.~2024ZD0607000); the Basic Research Program of Jiangsu (Grant Nos.~BK20233001, BK20241253, and BK20253009); the Jiangsu Funding Program for Excellent Postdoctoral Talent (Grant Nos.~2025ZB440 and 2025ZB852); the China Postdoctoral Science Foundation (Grant No.~2025M773331); the Postdoctoral Fellowship Program of CPSF (Grant No.~GZC20252202); the Fundamental and Interdisciplinary Disciplines Breakthrough Plan of the Ministry of Education of China (Grant No.~JYB2025XDXM413); the Science Challenge Project (Grant No.~TZ2025013); the AI \& AI for Science Program of Nanjing University; the Artificial Intelligence and Quantum Physics (AIQ) Program of Nanjing University; the Nanjing University PhD Student Zhujian Program; and the Fundamental Research Funds for the Central Universities.
The calculations were carried out using the supercomputers at the High-Performance Computing Center of the Collaborative Innovation Center of Advanced Microstructures and the high-performance supercomputing center of Nanjing University.

\FloatBarrier
\bibliographystyle{computsimcss}
\bibliography{references}

\clearpage
\appendix
\section{\label{app:computational_details}Computational details}

\subsection{Benchmark datasets}

The FeAl data originate from 21 fully relaxed 16-atom bcc Fe-Al supercells with Al concentrations from 0\% to 50\% and different atomic arrangements~\cite{Kotykhov2023ConstrainedDFTFeAl,Kotykhov2024MagneticForcesMTP}.
Additional configurations were generated by perturbing atomic positions, lattice vectors, and magnetic moments and by molecular dynamics at 300 K.
About 80\% of these additional configurations are perturbations and the remainder are molecular-dynamics snapshots; approximately 85\% of the configurations in the complete dataset contain nonequilibrium magnetic moments and therefore nonzero magnetic forces.

The noncollinear Fe benchmark uses the publicly released 16-atom LCAO subset of the Fe-DeepSpin dataset~\cite{Zheng2026FeDeepSpin}.
Its source workflow combines noncollinear spin-constrained DFT in ABACUS, active learning, and spin-lattice-dynamics sampling of bcc and fcc Fe.
The collinear Fe energy--volume dataset instead contains constrained ideal bcc and fcc structures spanning ferromagnetic (FM), antiferromagnetic (AFM), and double-layer antiferromagnetic (DAFM) orders, atomic volumes from 8 to 16~\AA\(^3\)/atom, and signed magnetic moments along the \(z\) axis.
Its stored \(x\) and \(y\) magnetic-force labels are identically zero.

The H-phase monolayer VSe\(_2\) data comprise 200 randomly displaced \(2\times2\times1\) supercells, 100 randomly displaced \(3\times3\times1\) supercells, and 121 fixed-structure six-atom cells scanning the orientations of the two vanadium spins.
The 421 configurations are used for both fitting and aggregate error evaluation, so the reported aggregate errors are in-sample statistics.
Three additional six-atom configurations evaluate ferromagnetic spins along \(z\), antiferromagnetic spins, and ferromagnetic spins along \(x\); the two vanadium atoms in the cell define the reported per-vanadium normalization.

\subsection{Model and training settings}

All four models are trained independently within the same general framework.
They use 64-dimensional embedding and hidden channels, truncate the structural, magnetic, and output tensor ranks at two, and use a maximum body order of three.
The structural radial filters contain eight Bessel functions from 1.0~\AA{} to the dataset-specific cutoff, followed by a two-layer \((64,64)\) multilayer perceptron with SiLU activation; a fifth-order polynomial envelope imposes the cutoff.

The FeAl model uses a 5.0~\AA{} cutoff, three structural and three spin interaction layers, and species-dependent linear-Chebyshev spin-radial intervals of 0.001--4.0~\(\mu_{\mathrm B}\) for Fe and 0--0.3~\(\mu_{\mathrm B}\) for Al.
The noncollinear and collinear Fe models use the same cutoff and layer counts, with a spin-radial interval of 0.001--5.0~\(\mu_{\mathrm B}\).
These three models include the zeroth-order spin-pair basis and a two-layer \((64,64)\) spin self-interaction head, and are trained without explicit SOC.
The H-phase monolayer VSe\(_2\) model uses a 6.0~\AA{} cutoff, a 1.5~\(\mu_{\mathrm B}\) spin-radial upper bound, two structural interaction layers, three spin interaction layers, and explicit SOC coupling.
The FeAl loss assigns equal unit weights to energy, atomic-force, stress, and magnetic-force MSEs, whereas the other three losses assign equal unit weights to energy, atomic-force, and magnetic-force MSEs.
Structure batch sizes are 16 for FeAl and VSe\(_2\) and four for both Fe models.
All models use AdamW with AMSGrad, zero weight decay, an initial learning rate of \(10^{-3}\), and 100 warmup steps; ReduceLROnPlateau uses a reduction factor of 0.8 for FeAl and VSe\(_2\) and 0.9 for the Fe models, with a patience of ten evaluations.
For the FeAl comparison, the reported mMTP row corresponds to the \(N_{\psi}=4\) magnetic-basis model fitted with \(w_e=1\), \(w_f=0.01~\text{\AA}^2\), \(w_s=0.001\), and magnetic-force weight \(w_t=0.1~\mu_{\mathrm B}^2\)~\cite{Kotykhov2024MagneticForcesMTP}.

\subsection{First-principles settings}

For FeAl, reference energies, atomic forces, stresses, and magnetic forces were computed using DFT and constrained DFT in ABINIT with PAW pseudopotentials, the PBE functional, a \(6\times6\times6\) \(k\)-point grid, and a 25 Hartree cutoff~\cite{Kotykhov2023ConstrainedDFTFeAl,Kotykhov2024MagneticForcesMTP}.

For noncollinear Fe, the released LCAO labels were calculated with the DZP-7au-v2.0 numerical atomic-orbital basis and contain energies, atomic forces, and magnetic forces obtained as energy derivatives with respect to the constrained atomic magnetic moments~\cite{Zheng2026FeDeepSpin}.
The first-principles calculations underlying the noncollinear Fe-DeepSpin benchmark included SOC; the source study noted that SOC effects in pure Fe are weak and smaller than the intrinsic errors of its model~\cite{Zheng2026FeDeepSpin}.

For the collinear Fe energy--volume benchmark, the reference labels were calculated with ABACUS in collinear LCAO mode~\cite{Chen2010ABACUSBasis,Li2016ABACUS} using the PBE functional~\cite{Perdew1996PBE}, an ONCVPSP norm-conserving Fe pseudopotential (\texttt{Fe.upf}), and the DZP-7au-v2.0 numerical atomic-orbital file \texttt{Fe\_gga\_7au\_100Ry\_4s2p2d1f.orb}.
The calculations used \texttt{ecutwfc=100}~Ry, \texttt{kspacing=0.14}~Bohr\(^{-1}\), Gaussian smearing with a width of 0.01~Ry, Broyden mixing, and \texttt{scf\_thr=1e-6}~Ry.
Signed site moments along the collinear \(z\) axis were constrained following the Lagrange-multiplier formulation of Dederichs \textit{et al.}~\cite{Dederichs1984ConstrainedSystems}, as implemented in ABACUS DeltaSpin~\cite{Cai2023DeltaSpin}, using \texttt{nspin=2}, \texttt{sc\_mag\_switch=1}, signed per-atom \texttt{sc} targets, \texttt{onsite\_radius=3.0}~Bohr, \texttt{sc\_thr=1e-7}~\(\mu_{\mathrm B}\), \texttt{nsc=100}, \texttt{alpha\_trial=0.01}, and \texttt{sccut=3.0}.

The H-phase monolayer VSe\(_2\) first-principles reference calculations were performed with ABACUS in the linear-combination-of-atomic-orbitals mode~\cite{Chen2010ABACUSBasis,Li2016ABACUS}.
They used the GGA/PBE exchange-correlation functional~\cite{Perdew1996PBE}, fully relativistic norm-conserving PseudoDojo pseudopotentials~\cite{VanSetten2018PseudoDojo}, double-\(\zeta\) plus polarization numerical atomic orbitals, a \(5\times5\times1\) \(k\)-point mesh, a 20~\AA{} out-of-plane cell length, and \texttt{ecutwfc=120}~Ry.
The calculations were noncollinear with SOC enabled and used constrained-spin settings following the DeltaSpin approach~\cite{Cai2023DeltaSpin}, with \texttt{sc\_mag\_switch=1}, \texttt{sc\_thr=1e-5}~\(\mu_{\mathrm B}\), \texttt{nsc=400}, \texttt{nsc\_min=2}, \texttt{alpha\_trial=0.01}, and \texttt{sccut=3}.
Gaussian smearing with a width of 0.01~Ry, Broyden mixing, a self-consistent-field threshold of \(10^{-6}\)~Ry, and a maximum of 400 SCF iterations were used; symmetry was disabled, atomic forces were evaluated, and DFT-D3 dispersion corrections were included~\cite{Grimme2010DFTD3}.

\section{\label{app:spin_dynamics}Finite-temperature spin-dynamics protocol}

The reported finite-temperature calculations propagate fixed-length spins while holding the atomic positions fixed.
Temperature points were sampled from 200 to 600 K, with a denser grid between 400 and 450 K.
Most temperatures were run for 50000 steps with a 1 fs time step; the 410--450 K region was extended at the same target temperatures because of the large fluctuations near the ordering crossover.
The notation and energy-to-field conversion follow the definitions in the main text; we use Gilbert damping \(\alpha=0.1\).
No explicit radial projection is applied to \(b_i^{\mathrm{ML}}\); the cross products remove its component parallel to \(s_i\).

The stochastic equation is interpreted in the Stratonovich sense and integrated using the semi-implicit B (SIB) method of Mentink \textit{et al.}~\cite{Mentink2010SIBLL}.
For a time step \(\Delta t\), independent vectors \(\eta_i\sim\mathcal N(0,I_3)\) define the dimensionless rotation increment
\begin{align}
  \Omega_i ={}&
  \frac{\gamma\Delta t}{(1+\alpha^2)\mu_i\mu_{\mathrm B}}
  \left[
    b_i^{\mathrm{ML}}
    +
    \alpha s_i\times b_i^{\mathrm{ML}}
  \right]
  \nonumber\\
  &+
  \frac{1}{1+\alpha^2}
  \sqrt{
    \frac{2\alpha\gamma k_{\mathrm B}T\Delta t}
    {\mu_i\mu_{\mathrm B}}
  }\,\eta_i .
  \label{eq:sib_increment}
\end{align}
The discrete thermal increments therefore obey
\begin{equation}
  \left\langle
    \Delta\Omega^{\mathrm{th}}_{i\lambda}
    \Delta\Omega^{\mathrm{th}}_{j\kappa}
  \right\rangle
  =
  \frac{2\alpha\gamma k_{\mathrm B}T\Delta t}
  {(1+\alpha^2)^2\mu_i\mu_{\mathrm B}}
  \delta_{ij}\delta_{\lambda\kappa}.
  \label{eq:sib_noise_covariance}
\end{equation}
The SIB update solves
\begin{equation}
  s_i' - s_i
  =
  -\frac{s_i+s_i'}{2}\times\Omega_i,
  \label{eq:sib_update}
\end{equation}
first for a predictor and then for the final spin after recomputing the learned spin force at the predictor midpoint.
The same Gaussian vector \(\eta_i\) is used in both stages.
This Cayley-type update preserves \(|s_i|\) algebraically, so no separate spin renormalization is applied after a time step.

The simulated system contains 900 vanadium sites with \(\mu_i=1\) and 1800 nonmagnetic selenium sites with zero stored moments; the selenium sites are not propagated by the spin update or included in the denominator of the normalized magnetic order parameter.
The initial spin state is the spin configuration stored in the input trajectory.
Observables are recorded every 100 integration steps, corresponding to 0.1 ps.
For temperatures sampled with continuation runs, the selected logs are the continuation segments initialized from the preceding trajectory.
At the standard temperature points, the first 5 ps of the selected log is discarded as equilibration, leaving 45 ps for production averaging.
For the 410--450 K continuation range, the first 20 ps is discarded and the longer remaining segment is used for production averaging, leaving approximately 200 ps of production data in the extended temperature window.
The remaining samples are divided into non-overlapping blocks of 200 recorded samples, or 20 ps per block, with incomplete trailing samples ignored.
Each standard-temperature trajectory therefore contributes two complete blocks, and its incomplete trailing 5 ps is discarded.
Error bars are block-to-block standard deviations.

\end{document}